\documentclass[12pt, preprint]{article} 
\usepackage{setspace}
\usepackage{graphicx}
\usepackage{dcolumn}
\usepackage{bm}
\usepackage{multirow}
\usepackage{amsmath}
\usepackage{amsfonts}

\title{Photoelectric charging of dust grains in the environment of Young Stellar Objects}
\author{Andreas Pedersen\\
\small{Instituto de Matem\'{a}tica Interdisciplinar, Fac. de CC Matem\'{a}ticas,}\\
\small{Universidad Complutense, 28040 Madrid, Spain}\\
\small{School of Science and Engineering, Reykjavik University, 101 Reykjav\'{i}k, Iceland}\\
\small{Science Institute, University of Iceland, 107 Reykjav\'{i}k, Iceland}\\
\\
Ana Ines G\'omez de Castro \\
\small{S.D. Astronom\'ia y Geodesia and Instituto de Matem\'{a}tica Interdisciplinar} \\
\small{Fac. de CC Matem\'{a}ticas, Universidad Complutense, 28040 Madrid, Spain}}

\begin{document}
\maketitle

\begin{abstract}
The evolution of disks around Young Stellar Objects (YSOs) is deeply affected by the YSOs ultraviolet (UV) radiation field especially in the 500-1100~\AA\ spectral range. This two dominant processes are; the photo-dissociation of H$_2$ molecules in the Werner and Lyman bands, and the emission of photo-electrons from dust grains when high energy photons are absorbed. Photo-electrons are an important source of gas heating. In this letter, dust grain charging when exposed to various possible UV fields in the YSOs environment is investigated.  Numerical simulation of the evolution of photo-electrons in the electric field created by the charged dust grains are carried out to obtain the charging profile of dust grains. From the simulations it appears that the different spectra produce significant quantitative and qualitative different charging processes.  Both the UV background and the Ae-Herbig star radiation field produce a relatively slow charging of dust grains due to the low fraction of sufficiently energetic photons. The radiation field of T Tauri Stars (TTSs) is harder due to the release of magnetic energy  in the dense magnetospheric environment. These numerical results have been used to propose a new simple analytical model for grain charging in the atmosphere of protostellar disks around TTSs susceptible to be used in any disk modeling.
It has been found that the yield decreases exponentially with the dust charge and that two
populations of photoelectrons are produced: a low energy population with mean kinetic energy $E= 2.5$ eV and a high energy population with $E=5.5-6$~eV; the energy dispersion within the populations is $\sim 1.3$~eV ($T\sim 1.5\times 10^4$~K). The high energy population is susceptible of dissociating the H$_2$  and ionizing some low ionization potential species, such as the Mg. These results add an additional role to dust on the chemistry of the layers just below the H$_2$ photoionization front. 
This photoelectic yield has been applied to a simple evaluation of the dust charge in the atmospheres of accretion disks ($\alpha$-disks).

\end{abstract}

\section{Introduction}

UV radiation strongly affects matter and the circumstellar environments around Young Stellar Objects (YSOs) are very sensitive to it. Of special interest is the interaction between the stellar UV radiation field and the accretion disk that channels mass infall onto the star.

In the diffuse atmosphere of the disk, the dominant interaction is the absorption of UV photons by the H$_2$ molecule in the Lyman and Werner bands; about 15\% of the photons cause photo-dissotiation of the rather inert H$_2$ molecule \cite{draine96} leading to both the formation of a photo-ionization front that propagates inwards the disk and an outward photo-evaporative flow on the surface, which carves the disk and produces a slow thermal wind \cite{HollenbachGorti2009} that has a fundamental role in disk evolution \cite{Alexander2006}. Indirect evidences of the absorption of UV radiation in young planetary disks come from the detection of H$_2$ electronic transitions that are pumped by the Lyman~$\alpha$ radiation from the star~\cite{herczeg02}.

Though H$_2$ by far is the main component of molecular clouds and protostellar disks ---the gas to dust mass ratio is $\sim 100$ and CO/H$_2 \sim 7 \times 10^{-5}$--- dust plays a relevant role as source of extinction and free electrons, not to mention its role in disk chemistry. The largest cross-section of a dust grain arises when interacting with radiation at wavelenghts $\sim 750$~\AA\ (declining both towards X-rays and optical wavelengths) thus, dust is more opaque to the optical/ultraviolet stellar spectrum than to the infrared radiation produced by itself.  
Furthermore does the interaction between dust grains and the UV radiation field result in the generation of a population of high energy photo-electrons. There is evidence of such hot electrons co-existing with molecular hydrogen around T Tauri stars (TTSs) \cite{ingleby09} and in the shocks of YSOs outflows with the environment, like in the Herbig Haro Object (HHO) HH~2. As an example, the observed H$_2$ emission spectrum of HH~2H can be fitted by radiation from H$_2$ at 1000~K that is collisionally excited by 20~eV electrons~\cite{raymond97}. There are, however, two problems with this interpretation. Firstly, it is unclear how hydrogen molecules can mix with hot electrons and radiate without dissociating.  Secondly, the H$_2$ temperature must be kept above $\sim 1000$~K to excite the vibrational levels and below $\sim 10,000$~K to avoid dissociation. Photo-electrons from UV irradiated dust could provide the needed thermostat, given that they are sufficiently energetic. 
Heating of the gas by collisions with photo-electrons depends on their emission rate, $J_{e}$, and their kinetic energy through:
\begin{align}
\Gamma = 4 n_H \sigma \int _{E _{min}}^{E_{max}} (E-eU) J_e dE
\end{align}
where $(E-eU)$ is the kinetic energy of an escaping photo-electron, $E$ is the energy of the impinging photons and $U$ the barrier to escape the grain potential, $n_H$ and $\sigma$ are the number density of H$_2$ and the collision cross section of H$_2$ molecules with the photo-electrons, in the simplest version of a dust grains fluid embedded in molecular gas. It should be noted that the kinetic energy of the photo-electrons depends on their interaction with the electric field generated by the charged electric grains and it is directly related with the grains charging profile.

In this paper, the results from detailed numerical simulations of the charging process of dust grains are reported, for various UV radiation fields of interest for star formation research (photoelectric effects by X-rays and Auger effects are beyond the scope of this paper) Also, a simple parametrization of $J_e$, the photoelectric current produced by a dust grain, is derived for grains submitted to the irradiation of the magnetosphere of a TTS. The dependence of the
photolectric yield on the dust grain charge is analyzed in Sect.~4 and the population of ejected photo-electrons is characterized in terms of it kinetic energy. Finally, the relevance of photoelectric charging of dust in the atmospheres of $\alpha$-disks is discussed.

\section{Numerical simulations}

The applied scheme to account for the interaction between incoming photons and dust grains follows four steps:
\begin{enumerate}
\item Capture of ionizing photons by the dust grain.
\item Yield of photo-electrons released from the surface of the grain-core.
\item Determination of the initial kinetic energy of a released photo-electron.
\item Dynamics of the electrons in the local electric field that results from the charged grain-core and the surrounding electronic cloud.
\end{enumerate}
The first three phases are based on the scheme originally proposed by~\cite{draine78} for the treatment of the interaction between radiation and dust grains. All interactions are parametrized for the most common silicate dust grains (see~\cite{pollack94, sargent09}).

\subsection{Photon capture}
For a photon with the wavelength $\lambda$,  the cross section for capture/absorption is given by:
\begin{align}
\label{equ:ImpingingPhotons}
\sigma(r_{grain}/\lambda,\Im(\tilde{n})) = Q_{abs}(r_{grain}/\lambda,\Im(\tilde{n})) \pi r_{grain}^2
\end{align}
where $\tilde{n}$ is the complex refractive index for the grain material and $r_{grain}/\lambda$ is the ratio between the grain size and the wavelength of the impinging photon.  The absorption efficiency $Q_{abs}$ is obtained by applying the MIE-theory using the functional form of the refraction index for silicate as stated by~\cite{draine03}.

\subsection{Yield of photo-electrons}
\begin{figure}
\includegraphics[width=8.5cm]{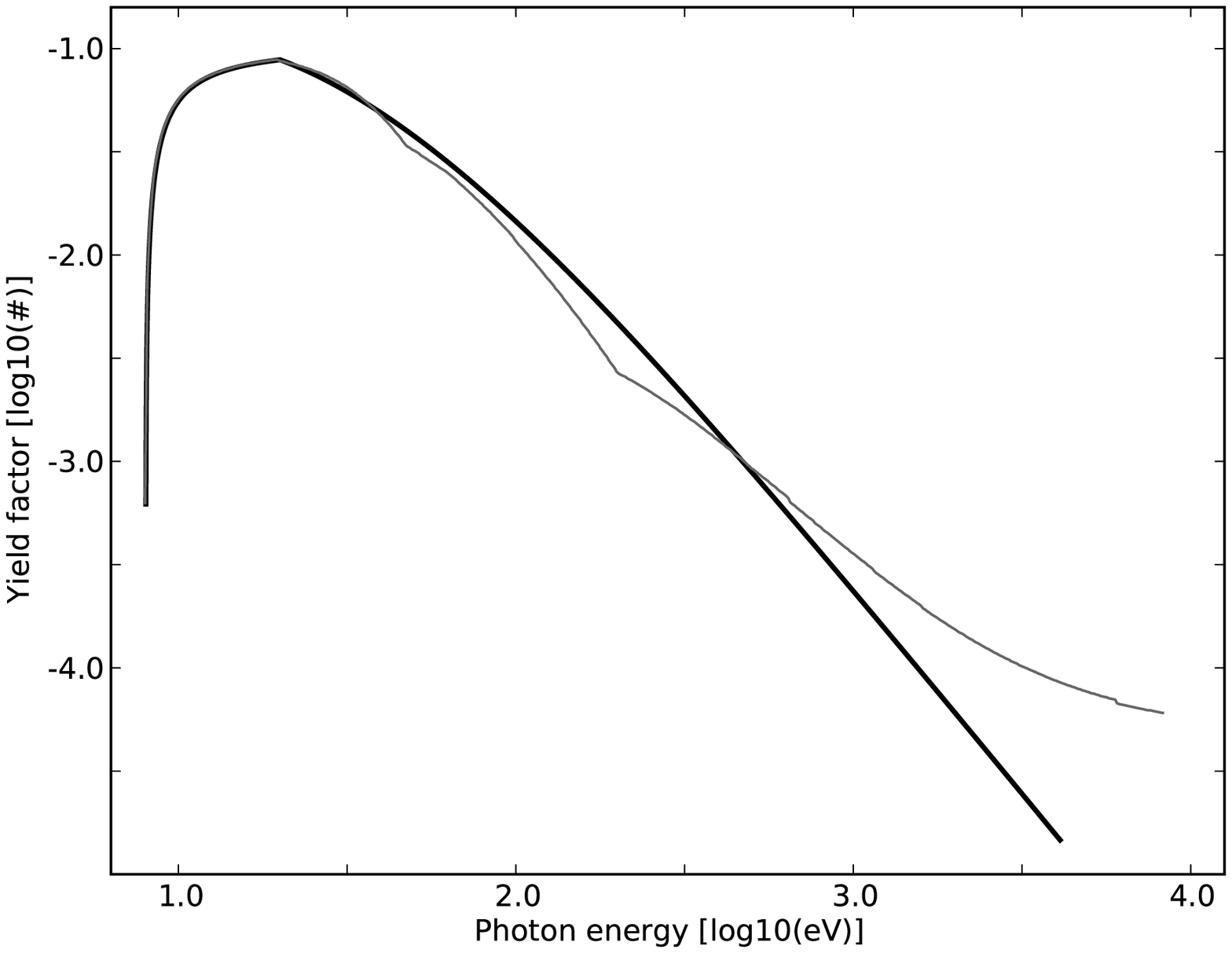}
\caption{Yield as a function of photon energy. Grey curve is model by~\cite{weingartner06} for the decrease at high photon energies, bold black curve is the new proposed approximate model, Equ.~\ref{equ:YieldAtomic}.
}
\label{fig1}
\end{figure}

For a bulk sample the electron yield produced by a photon, $Y$, is approximated following the scheme by~\cite{weingartner01}:
\begin{align}
\label{equ:YieldBulkWeingartner}
Y(\Theta) = \frac{\alpha(\Theta/\chi_{mol})}{1+\beta(\Theta/\chi_{mol})}
\end{align}
where  $\Theta = h\nu -\chi_{mol}$, and the parameters $\chi_{mol} = 8~eV$, $\alpha = 0.5$ and $\beta = 5$ represent silicates. For this yield model to be valid the impinging photon must interact with a molecular orbital and its energy must be  $< \chi_{atom}$. For the more energetic photons a decrease in yield occurs since they interact with the atomic orbitals, which results in a weaker coupling. To account for this behavior we suggest a simple and analytic approximation:
\begin{align}
\label{equ:YieldAtomic}
Y(\Upsilon)=\left(\frac{\Upsilon-a}{a}\right)^{b}
\end{align}
%
where $\Upsilon = h\nu - \chi_{atom}$ 
whereas $a$ and $b$ are fitting parameters.  The values used for silicates are:  $\chi_{atom} = a = 20$~eV and  $b = -2$.  For these values a good fit to the more accurate model by~\cite{weingartner06} is obtained, see Fig.~\ref{fig1}. With this additional term the full expression for the yield of photo-electrons from a bulk sample 
becomes:
\begin{align}
\label{equ:YieldBulk}
Y(h\nu)=
  \begin{cases}
  \frac{\alpha(\Theta/\chi_{mol})}{1+\beta(\Theta/\chi_{mol})}					& \text{ if \: $\chi_{mol} < h\nu \leq \chi_{atom}$} \\
  \frac{\alpha(\Theta/\chi_{mol})}{1+\beta(\Theta/\chi_{mol})} \left(\frac{\Upsilon-a}{a}\right)^{b}	& \text{ if \: $\chi_{atom} < h\nu$}
\end{cases}
\end{align}
Size effects result in a dependency between yield and the curvature of the grain surface as an increased curvature results in a stronger electric field within the grain-core. Moreover, the distance a photo-electron has to travel within the grain ---to reach the surface--- decreases. These effects have been modeled accurately by~\cite{watson72} however, the approximate model as suggested by~\cite{draine78} has been implemented into our numerical code:
\begin{align}
\label{equ:GrainEnhance}
Y(r_{grain})=\frac{\beta}{\alpha}^2 \frac{\alpha^2 -2\alpha+2-2\exp[-\alpha]}{\beta^2 -2\beta+2-2\exp[-\beta]},
\end{align}
where $\beta=r_{grain}/l_{photon}$ is the inverse penetration depth of the photon, $\alpha=r_{grain}/l_{photon}+r_{grain}/l_{electron}$ relates the penetration depth of the photon, $l_{photon}$, and the length for a photo-electron to travel within the grain-core, $l_{electron}$. This distance is assumed to be $l_{electron}=10 \text{~\AA}$ and represents the distance from the molecular orbitals containing loosely bound electrons to the surface of the grain. The photon penetration depth is given by  $l_{photon}=\lambda[4\pi \Im(\tilde{n})]^{-1}$, where $\lambda$ is the wavelength of the impinging photon and $\Im(\tilde{n})$ is the imaginary part of the refractive index of the grain material. The total yield is then given by:
\begin{align}
\label{equ:ElectronRelease}
Y_{tot} &= Y(h\nu)Y(r_{grain})
\end{align}
Photo-electrons with a total yield exceeding unity are always be released. When the yield is less than unity a Monte Carlo approach is applied and a photo-electron only gets released if the determined yield exceeds a random number drawn from a normalized even distribution.

\subsection{Initial kinetic energy of the released photo-electrons}

The distribution function for the initial kinetic energy of photo-electrons is assumed to be parabolically shaped as suggested by~\cite{weingartner01}
and a normalized distribution might be given by:
\begin{align}
\label{equ:Shifted}
f(E) &= \frac{-6 E^{2} + 6 E E_{high}^{*}}{E_{high}^{*3}}
\end{align}
where $E^{*}_{high}$ is the excess energy of the impinging photon when the ionization energy has been subtracted i.e., the highest possible kinetic energy a released photo-electron might get. 

As the distribution is normalized
for the interval $[0, E^{*}_{high}]$ a random number, $\mu$, between zero and one is used to set the kinetic energy, $E_{\text{kin}}$, for the released photo-electron through the expression:
\begin{align}
\mu &= \int_{0}^{E_{\text{kin}}}\frac{-6 E^{2} + 6 E E^{*}_{high}} {E_{high}^{*3}}dE\\
E_{\text{kin}} &= \text{Root}(-6E^{3}/(3E_{high}^{*3}) + 6E^{2}/(2E_{high}^{*2})^{1/2}- \mu ) 
\end{align}
where Root() is the root of the polynomial within the above stated energy interval.  

\subsection{Dynamics of electrons in the electric field of a charged dust grain}

In the event that a photo-electron is released, a stationary positive charge is assigned to the center of the dust grain, which is assumed to be a spherical core with the radius  $R$. This radius is furthermore used to determine if an orbiting electron recombines with the charged grain-core, which is the case if it gets within a distance of $R$ from the center of the grain.

The electron emission site from the grain surface is selected randomly on the hemisphere facing the source. The site is obtained from a three component vector drawn from a Gaussian distribution scaled by the core radius.  Under the assumption that the photon-source is placed at $x = \infty$, it is only the half-sphere where $x$ is positive that undergoes emission processes and the site vector becomes $[\text{Abs}(\hat{r_x}), \hat{r_y}, \hat{r_z}]$. Besides defining the emission site the vector also represents the surface normal ---the tangential plane at the emission site--- used to determine the half-sphere, which the direction of the initial velocity vector $[\hat{v_x}, \hat{v_y}, \hat{v_z}]$ is restricted to.

While an electron orbits a charged grain-core it might escape the local environment that occurs if the boundary of the simulation cell is reached; a cube of side length 100 $\mu m$. At this point,  is the background electric field assumed to become dominant over the local field.  To simulate the trajectory of the released electrons a Velocity Verlet algorithm (see Appendix A) and an accurate Coulomb potential are applied; the time step is $2.5 \times 10^{-15}$~s.  The grain-core is kept fixed in the center of the simulation cell.

\section{Dust grains interaction with YSOs radiation fields}

Four irradiation fields have been selected for this study. Two of them are black bodies at temperatures of 50,000~K  and 10,000~K  simulating the photospheric radiation fields from O and A stars, respectively. A third spectrum is the fit to the interstellar radiation field made by  \cite{draine78} based on the  early measurements of the Copernicus and Voyager missions~\cite{habing68}. This soft spectrum is used as a standard in chemical modeling of the disk; often the intensity of the radiation field is scaled with respect to the mean interstellar radiation field by means of the standard $G$ factor (typically $G=10^4$ at 100 AU i.e. \cite{semenovetal2004}). Finally, a {\it hard} UV spectrum has been built for this study to {\it model} the hard energy radiative output of the TTSs. Notice that there is little information on the spectral energy distribution of the TTSs in the range 10~\AA\ $\leq \lambda \leq$ 950~\AA . Furthermore, only few TTSs (TW~Hya, RU~Lup and T~Tau) have been observed in the 950-1100~\AA\ range with the Far Ultraviolet Spectroscopic Explorer~\cite{wilkinson02, herczeg02}. There are however, strong evidences of energetic processes ongoing in the atmospheric/magnetospheric environment of the TTSs. On the one hand, the near UV (1150~\AA\ $\leq \lambda \leq$ 3200~\AA) indicates atmospheric structures extending to the inner border of the gas disk (see~\cite{gomez09} for a review). On the other hand, the observed X-ray luminosity of YSOs is on average about $10^{30}$erg~s$^{-1}$ and the energy distribution is well fitted by the thermal radiation of optically thin plasmas at temperatures of $\sim$5-30~MK (see i.e.~\cite{guedel08}), for comparison the Solar X-ray luminosity is about 10$^{27}$erg~s$^{-1}$ and the plasma temperatures 1-2~MK. Thus, the propagation of the X-ray radiation on the dense ($\geq 10^9$~cm$^{-3}$) and extended (2-4 R$_*$) atmospheres/magnetospheres of the TTSs is expected to produce a rich and energetic spectrum. The spectrum used in this work has been used to fit the observed CIV/Si~III] and C~III]/Si~III] lines ratios~\cite{gomez03}. 

\begin{figure}
\includegraphics[width=8.5cm]{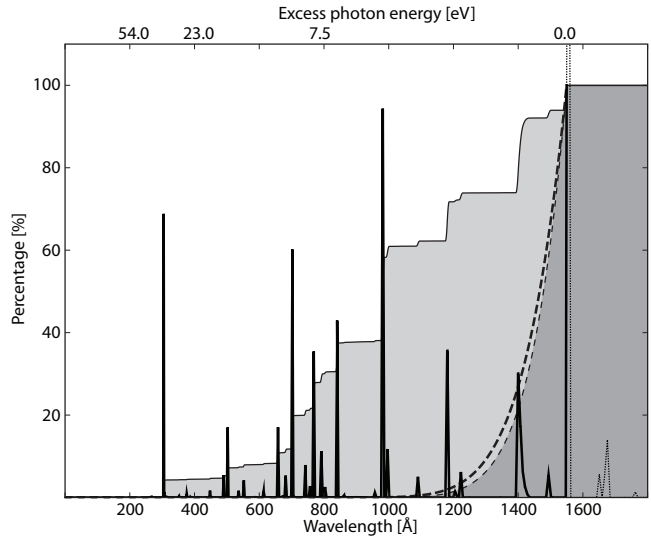}
\caption{Distribution functions of the photon counts  and the cumulative value for ionizing radiation.  Solid line is the spectrum for a T Tauri star and the dashed line is the BB-50kK that both have been normalized to enclose an unit area.  Grey shaded areas are the cumulative percentage of the spectra below the specific wavelength.
}
\label{fig2}
\end{figure}

The spectral energy distributions (in photons~$ {\rm s}^{-1}{\rm cm}^{-2}$) have been normalized to the same bolometric flux. The ionizing part of the TTSs model spectrum and the black body at 50,000~K are plotted in Fig.~\ref{fig2}; also the cumulative distributions of ionizing photons are plotted for reference. Notice that all the energy distributions are smooth besides the TTS distribution. 

\begin{figure}
\includegraphics[width=8.5cm]{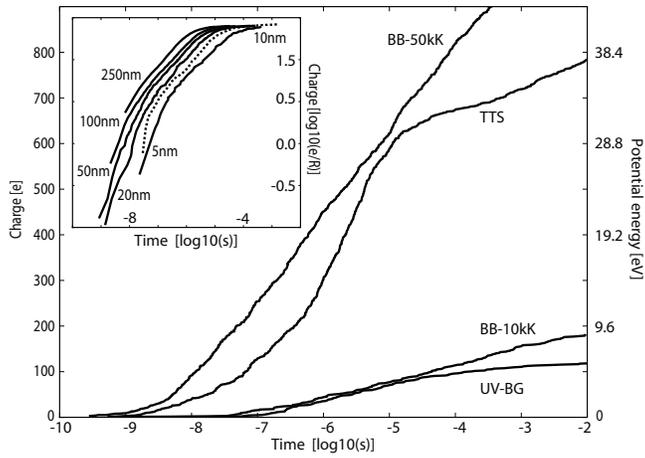}
\caption{Charging profiles for a 30~nm grain when exposed to a spectrum from a T Tauri star (TTS), a Blackbody spectrum at a temperature of either 10,000~K (BB-10kK) or 50,000~K (BB-50kK), and the UV background (UV-BG).  It should be noticed that for the BB spectra linear profiles are obtained whereas the TTS results in a S-shaped profile and the UV-BG saturates at a plateau level.  Inset, charging profiles normalized with the grain radius for different grain sizes when exposed to the TTS spectrum.
}
\label{fig3}
\end{figure}

The simulated charging profiles of a 30~nm dust grain are shown in Fig.~\ref{fig3} for the four energy distributions: black body at 50,000~K (BB-50kK), black body at 10,000~K (BB-10kK), UV background (UV-BG) and TTS. From the simulated results, it appears that the TTS spectrum gives raise to a S-shaped profile whereas linear profiles are obtained for the two black body spectra.  The UV-BG saturates at a charge of approximately 125~$e$.  It also appears that the quantitative charging when exposed to the TTS or the BB-50kK is alike and that the UV-BG or BB-10kK gives raise to the similar charging rates. The relative slow charging obtained with the UV-BG or BB-10kK is a consequence of a low ratio of photons sufficiently energetic to both ionize an atom and provide the needed kinetic energy for this released electron to escape the charged grain-core.  Whereas the relative fast charging observed for the TTS or BB-50kK is possible because of a larger ratio of energetic photons.

The saturation threshold apparent in the figure for the UV-BG spectrum is caused by the high energy cut-off of the energy distribution. The shortest wavelength corresponds to a photon energy of 14~eV that gives an excess energy of 6~eV, which equals the needed kinetic energy an electron requires to escape the electric field created by a grain of charge 125~$e$ and a radius of 30~nm that explains the observed plateau level. The linear profiles for the BB spectra are consistent with the experimental observation by~\cite{sickafoose00} and is a behavior that is accounted for by theoretical models for the charging of a capacitor.  From the linearity of the charging profiles it can be concluded that effects caused by the interactions between released electrons are negligible for the simulated photon fluxes. The S-shaped charging profile caused by TTS is a consequence of the line structure of this spectrum. From the cumulative distributions it also appears that the ratio of photons at high energies is larger in the TTS spectrum than in BB-50kK, see Fig.~\ref{fig2}. A consequence of this shows in the interval from $[1~\mu s, 10~\mu s]$ in Fig.~\ref{fig3} where the TTS almost manages to reach the same charge state as obtained by the BB-50kK.  If the energy scale shown rightmost in Fig.~\ref{fig3} is compared with the one uppermost in Fig.~\ref{fig2} it emerges that the energy level at which the curvature changes after $10^{-5}$ s is coinciding with the first significant peak in the TTS spectrum.

\subsection{Dependence on the size of the dust grain}

A series of simulations have been carried out to address the dependency between the grain size and the charging profile.  Only the TTS spectrum is used for this purpose.  As expected, the maximum charge a grain may achieve increases as the grain size increases.  However, the maximal charge is determined by the radius of the grain rather than its volume, see inset in Fig.~\ref{fig3}.  As a result, the maximal charge-to-mass ratio increases as the grains becomes smaller.  This behavior is caused by the electric potential at the grain surface since the energy barrier an electron must overcome to escape the grain surface is $\propto Q/r_{\rm grain}$.

\section{Model of the photoelectric emission in the TTSs environment}
\label{sec:ChargingModel}

Two main results can be derived from the simulations: (1) the charging profile qualitatively remains unaffected when the grain size varies
and (2) the maximal charge obtainable by a grain is solely determined by its radius. Thus, to reproduce the charging profile of an arbitrary sized grain it is sufficient: (1)  to fit the general functional form of the charging profile for a grain and (2) multiply it by a scaling factor, which is a linear function of the radius.

\begin{figure}
\includegraphics[width=8.5cm]{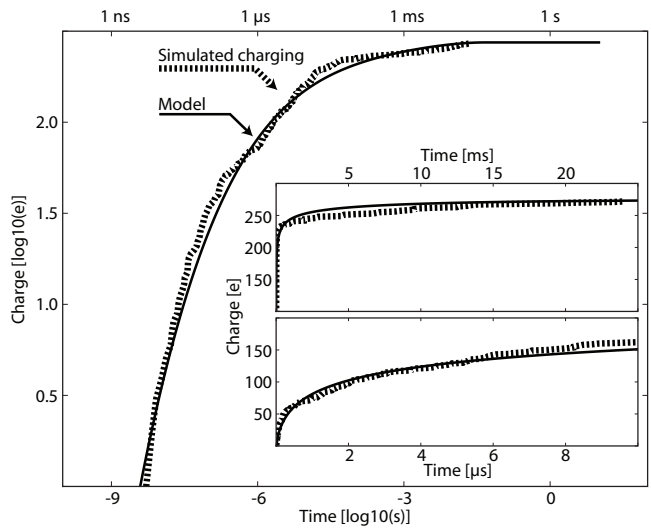}
\caption{
Charging model for an arbitrary sized grain.
The fourth-order polynomial used to model the dependence between time and the charge state of a grain shown in a double logarithmic plot togetether with simulated data, the parameters used are as stated in Equ.~\ref{equ:chargingModel}. It should be noted that the asymptotic plateau value used in the model is set to 275~$e$. Insets show the simulated data and the fitted model for the charge state as a function of time.
}
\label{fig4}
\end{figure}

The charging profile of a fiducial 10~nm dust grain has been fit. The grain is irradiated with photons from the TTS spectrum until it reaches a saturation threshold of 275~e. During the simulation, 1-2 electrons usually orbit the grain before they either escape or recombine with the grain core. Electrons leaving the computational box are lost to the background. The charging profile has been normalized to a photon rate, $S_{19}$, of 10$^{19}$~s$^{-1}$cm$^{-2}$ with energies above the dust grain threshold for photo-ionization that corresponds to photons with wavelengths $\lambda < 1550~$\AA. This photon flux is equivalent to what is expected from an extended magnetosphere of radius $3~R_{\odot}$ producing a total ionizing luminosity of 0.03~L$_{\odot}$. Notice that for the TTS spectrum, this photon rate is 4.8 times higher than the photon rate in the Lyman continuum ($1100$~\AA $< \lambda < 912$~\AA ). The charging profile is represented in Fig.~4 and the polynomial fit to it is given by,
\begin{align}
\label{equ:chargingModel}
\log_{10} Q(x) &= -0.0015 x^4 -0.00455 x^3 -0.0177 x^2 +0.0213 x + 2.4326 & \\
x &= \log_{10} (S_{19} t) =\log_{10} \left(\frac{S}{10^{19}~{\rm s}^{-1}{\rm cm}^{-2}} \frac {t}{{\rm 1 ~ \mu s}}\right)    \in [-8.4, -1.3]
\end{align}
The functional dependence for $Q(x,r_{grain})$ is then described by:
\begin{align}
\label{equ:chargingModel2}
Q(x,r_{grain} ) &= \frac{r_{grain}}{{\rm 10~nm}}
  \begin{cases}
  0 & \text{ if   \: $S_{19} t < 4 \times 10^{-9}$} \\
  Q(x) & \text{ if   \: $4 \times 10^{-9} \le S_{19} t \le 5\times 10^{-2}$} \\
  275 & \text{ if \: $\S_{19} t > 5 \times 10^{-2} $}
  \end{cases}
\end{align}
For this grain size the asymptotic value for the final charge is set to  275~e as only photons with wavelengths less than $\sim 258$~\AA\ will be able to cause any further charging that are assumed to be of an insignificant rate, see Fig.~\ref{fig2}. The charging profile for the proposed analytical expression and the simulated data are shown in Fig.~\ref{fig4}. From the figure it appears that the proposed model reproduces the main features of the numerical simulations being an abrupt initial charging and a final plateau level.  For the remaining intermediate interval a good agreement is obtained between the simple analytical model and the more accurate numerical results.

\subsection{The dependence of the photoelectric current on the grain charge}

To derive a simple parametrization of $J_e^o$ on the grain charge $Q$, a new series of simulations was conducted where the grain cores are kept at a constant charge level.  In these simulations the kinetic energy of electrons escaping the grain is recorded however, rather than loosing these electrons to the background, they are reassigned to the grain-core to maintain its charge state. To generate the data grains at different charge levels are exposed to the TTS spectrum for 1~$\mu s$. Exactly, the same series of photons is used for each run.
The number of ejected photo-electrons is found to decrease exponentially as, 
$\propto 10^{-0.016Q}$, 
where $Q$ is the charge of the dust grain and the current of photo-electrons, $J_{e}^e$, {\it emitted from the 10~nm grain}, might be expressed as, 
\begin{align}
\label{equ:fluxModel}
J_{e}^e = 6.3 \times 10^2 ~ \frac{\rm electrons}{\mu {\rm s}} S_{19} 10^{-0.016 Q}
\end{align}
Under the assumption that photo-electrons are emitted isotropically from the 
10~nm dust grain surface, the photoelectric flux is given by,
\begin{align}
\Phi _e^e &= 2 \times 10^{20} \frac{\text{electrons}}{\text{cm}^2\text{s}} S_{19}
 10^{-0.016 Q}
\end{align}
that can be scaled as,
\begin{align}
\Phi _e^e &= 2 \times 10^{20} \frac{\text{electrons}}{\text{cm}^2\text{s}} S_{19}
\left( \frac {r_{grain}}{10{\rm ~ nm}} \right) ^{-1}  10^{-0.016 Q}
\end{align}
for any arbitrary radius of dust grains, where the dependence of the dust grain charge
on the radius is inserted from Eq.\ref{equ:chargingModel2}.

\subsection{Energy distribution of the photo-electrons}

\begin{figure}
\includegraphics[width=8.5cm]{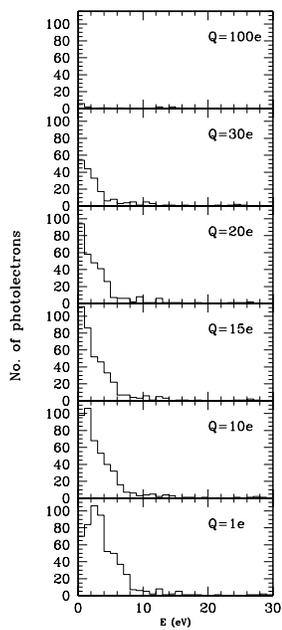}
\caption{
Energy distribution of the ejected photoelectrons. As the charge of the grain increases, only the high energy electrons are able to reach the boundary of the system and escape the grain-core.
}
\label{fig5}
\end{figure}

The energy distribution of the ejected photo-electrons follows a bimodal distribution with a narrow peak at about 2.5~eV and a broad shoulder peaking at $\sim 5.5-6$~eV. As the charge of the grain grows ---which results in an increased local electric field--- only electrons in the high energy tail of the distribution are able to escape and the electrons flux decreases dramatically, as shown in Fig.~\ref{fig5}. This distribution is caused by the energy distribution of the TTSs spectrum
irradiating the dust grain;  the first peak at 2.5 eV is produced by the Ly$\alpha$ continuum
while the shoulder at 5.8 eV is caused by high energy far UV photons produced by highly ionized species in the 700\AA\ - 850\AA\ wavelength range. A useful parametrization of this behavior is given by,
\begin{align}
\label{equ:energyDistribution}
\frac {dJ_e^e (Q,E)}{dE} &= 105 \exp \left( \frac{-(E'-2.5)^2}{3} \right) + 40 \exp \left( \frac{-(E'-5.8)^2}{4} \right) 
\end{align}
where $E' = E+0.14(Q-1)$ and $J_e^e(Q,E) $ is the number of photo-electrons ejected with energy in the range ($E-0.5$, $E+0.5$)~eV, for a grain with charge $Q$ and radius 10~nm.  The fraction of photo-electrons ejected with energies susceptible to ionize the hydrogen (13.6 eV) is 3\% for $Q=1$.
The parametrization of the photo-electrons energy distribution allows to address the effect of the UV irradiation, in terms of the two  electron populations produced:
\begin{itemize}
\item A low energy population with $E=2.5$~eV and dispersion 1.2~eV.
\item A high energy population with $E= 5.8$~eV and dispersion 1.4~eV.
\end{itemize}
where the dispersion is a measure of the temperature of the electron beam: $1.4 \times 10^4$~K and $1.6\times 10^4$~K for the low and high energy populations respectively. Notice that the high energy population is well above the dissociation threshold of the H$_2$ molecule (4.52~eV) and may contribute to the ionization of Mg ($\chi _i = 7.64$~eV), that is a fundamental charge carrier for protostellar disks \cite{OppenheimerDalgarno1974, IlgnerNelson2008}. 

\section{Discussion: dust charging in TTSs' disks}

The dust charging profile derived in Sect.~4, provides a useful parametrization of the photoelectric current generated by dust grains irradiated by a TTS spectrum. To determine the final charge of dust grains is a much more complex problem since other processes must be taken into account, including collisions with charged particles or charge transfer reactions between the dust grains and the surrounding molecules (see \cite{vanzadenhoffetal2003} \cite{semenovetal2004} for a detailed evaluation of the ionization degree and the main chemical processes in protostellar disks). 
For a generic electron density, $n_e$, and temperature, $T_e$, the final dust grain equilibrium charge is determined by the balance of the photoelectric emission current, $\Phi_e^e$, and the  driven into the positively charged grain by the collisions with the surrounding thermal electrons.
The flux of absorbed electrons per unit surface on the dust grain, $\Phi _e^a$, \cite{spitzer78}  is given by,
\begin{align}
\Phi _e^i &= \xi _e n_e \left( \frac {8 k T}{\pi m_e}\right)^{1/2} \left(1+\frac{e^2Q} {r_{grain}kT} \right) 
\end{align}
where $\xi _e$ is the sticking probability of the electrons striking the grain and $\left(\frac {8 k T}{\pi m_e}\right)^{1/2}$ is the thermal velocity of the electrons. For a 10~nm radius dust grain,
\begin{align}
\frac{e^2Q}{r_{grain}kT} &= 1.67 Q \frac {10^3 ~ {\rm K}}{T} 
\end{align}
and thus, 
\begin{align}
\Phi _e^a &= 6.4 \times 10^7 ~ \frac {\text{electrons}}{\text{cm}^2 \text{s}} \xi _e n_e 
\left(\frac {T}{10^3 ~ {\rm K}} \right) ^{1/2} 
\left(1+1.67  \frac {10^3 ~ {\rm K}}{T} Q \right)
\end{align}
Both, $\Phi _e^a$ and $\Phi _e^e$ depend on the environmental conditions. Let us make an estimate of the relevance of photoelectric charging in the atmospheres of TTSs' disks.

\subsection{Physical properties of $\alpha$-disks}

Assuming a simple $\alpha$-disk model \cite{ShakuraSunyaev} , the variation of the disk density
and temperature is readily prescribed in terms of $R$, the distance to the disk axis and $z$,
the height above the disk midplane for a given accretion rate, $\dot M_a$ and mass of the
central star, $M_*$. The density of the disk, $n$, is  given as:
\begin{align}
n(R,z) = n (R,0) \exp (-z/H(R))
\end{align}
with the density in the disk mid-plane being,
\begin{align}
n(R,0) = 9.02\times 10^{13} ~ {\rm cm}^{-3} \left( \frac {\dot M_a}{10^{-8} M_{\odot} {\rm yr ^{-1}}} \right) ^ {5/8}
\left(  \frac{M_*} {M_{\odot}} \right) ^{5/8} 
\left( \frac {R}{0.1 ~ {\rm AU}} \right) ^{-15/8} 
\end{align}
for $R >> R_*$, as it is the case, and assuming that the main component of the disk mass is molecular
hydrogen. The scale height of the disk, $H(R)$, is given by,
\begin{align}
H (R) = 3.1 \times 10^{10} ~ {\rm cm} \left( \frac {\dot M}{10^{-8} M_{\odot} {\rm yr ^{-1}}}\right) ^{1/8}
\left ( \frac{M_*} {M_{\odot}} \right) ^{-3/8} 
\left( \frac {R}{0.1 ~ {\rm AU}} \right) ^{9/8}
\end{align}
The disk temperature at a given radius, $R$, is derived from the $\alpha$-disk prescription to be,
\begin{align}
\frac {T} {10^3 K} = 0.48 \left( \frac {\dot M_a}{10^{-8} M_{\odot} {\rm yr ^{-1}}} \right) ^{1/4}
\left( \frac{M_*} {M_{\odot}} \right) ^{1/4} 
\left( \frac {R}{0.1 ~ {\rm AU}} \right) ^{-3/4}
\end{align}

\subsection{UV irradiation of $\alpha$-disks}

$S_{19}$ varies several orders of magnitude within the dust disk and depends on the distance to the star and the opacity of dust grains to the UV radiation. UV radiation penetrates through the disk atmosphere creating a photo dissociation region that has been extensively studied (see i.e. \cite{HollenbachGorti2009}). It is expected that the ionization front driven by the UV radiation splits the atmosphere into two regions, an upper ionized layer where H~I is fully dissociated and a transitional region where the ionization fraction of H~I decreases to negligible values and the electron density is controlled by Mg~I ionization. Further inside the disk
only X-ray photons are able to penetrate and ionize the dust grains. 

Accretion disks around TTSs do not reach the stellar surface; the stellar magnetic field disrupts the
inner area of the disk. Typically accretion disks are considered to extent from some 3-4 stellar radii to some 500~AU (see i.e. \cite{MillanGabet2007}) however, the inner most area
of the disk is devoided of dust grains. The UV radiation from the star melts the dust 
grains to a radius of $\sim 0.1-0.2 $~AU. This radius roughly corresponds to the silicates melting
radius \footnote{At thermal equilibrium, a dust grain irradiated by a TTS with luminosity $L_{*}$ is heated to
a temperature $T$, such that $ \pi r_{grain}^2 \frac {L_*}{4\pi R^2} = \frac {4} \pi r_{grain}^2 \sigma T^4$.
For $T$ about 2/3 of the silicates melting temperature ($T=2005$K), the value of
R derived is 0.1~AU, for a solar luminosity star.}. At this inner radius, 
$S_{19} = 0.073$ and decreases steadily to $S_{19} = 2.9 \times 10^{-9}$ at 
500~AU in the outskirts of the disk. 

The stellar flux is absorbed by the ionized recombining upper layer
and by dust absorption. The transfer of radiation the stellar UV flux, $F_{UV}$, in this layer is determined by,
\begin{align}
\dfrac {dF_{UV}(R,z)}{dz} = -\alpha _B n(R,z)^2-n(R,z)\sigma _{UV}F_{UV}(R,z)
\end{align}
where $\alpha _B$ is the case B hydrogen recombination coefficient that at 10$^4$K is 
$\alpha _B \simeq 2.53 \times 10^{-13}$~cm$^{3}$~s$^{-1}$ \cite{StoreyHummer},  $\sigma _ {UV}$ is 
the dust grains absorption cross section per hydrogen nucleus, typically $\sigma _{UV} = 0.61
\times 10^{-21}$~cm$^{-2}$ \cite{BertoldiDraine1996} and $n(R,z)$ is the hydrogen density
at cylindrical coordinates $(R,z)$ in the disk. The general solution to this equation is:
\begin{align}
F_{UV} (R,z) = \exp{ -\tau (z)} ( F_{EUV,0} - \int _{0} ^z dz' \alpha_B n^2(R,z') \exp{-\tau (z')}
\end{align}
with $\tau (z) = \sigma _ {UV} \int _0 ^z n(R,z') dz' = \sigma _{UV} N_H$ being $N_H$, the
column of H~I and $F_{UV,0}$ the radiation field at the disk surface. However, the stellar
radiation ought to cross the wind departing from the disk that may effectively shield the
disk against photoionization at early stages (see \cite{FFGdC} for an analysis of 
the effect caused by the fast centrifugally driven magnetic winds and \cite{HollenbachGorti2009}
for an analysis of the effect caused by slow, thermally driven winds). In such a case,
the dominant absorbing component is the recombining H~I; an optical depth of $1$ is reached 
for an H~I column of $10^{17}-10^{18}$~cm$^{-2}$. Moreover,  the properties of dust
grains in the disks atmospheres are ill determined but significant deviations from the
standard dust grains size distribution in the interstellar medium are 
expected\footnote{Following \cite{pollack94}, the dust particles size distribution in protostellar disks can be modelled,\\
$N_d (r_{\rm grain}) = N_{d,0} r_{\rm grain}^{-3.5}$, for  $r_{\rm grain} \in [0.005-1] \mu$m \\
$N_d (r_{\rm grain}) = N_{d,0} r_{\rm grain}^{-5.5}$, for  $r_{\rm grain} \in (1-5] \mu$m \\
According to this distribution, the average dust grain size is
$<R> = 0.008 \mu$m and the maximum charge that such grains can achieve is $<Q_{max}> = 220 e^-$.}; 
large dust grains are expected to be more numerous due to the high densities in the disk. 

In summary, in the absence of well determined properties of the absorption of the UV flux on the disk
atmosphere, $S_{19}$ can be parametrized as, $S_{19,0}\exp ^{-\tau}$, where $S_{19,0}$ is the
flux at the disk surface and $\exp ^{-\tau}$ parametrizes the extinction as a function of
an unknown optical depth, $\tau$.
Notice that the gas and dust columns are very high and that UV radiation will affect only the upper atmospheric layer. Typically, the dust grains absorption cross section per hydrogen nucleus,$\sigma _ {UV}$, is  $\sim 0.61 \times 10^{-21}$~cm$^{-2}$ \cite{BertoldiDraine1996}. 
Thus, in the innermost region of an $\alpha$-disk, the dust column reaches (1/$\sigma _{UV}$) 
already at $z \sim 5 H$  (for a standard accretion rate of 
$10^{-8} M_{\odot} {\rm yr ^{-1}}$).

\subsection{Relevance of photoelectric charging in the atmosphere of an $\alpha$-disk}

Some hints on the relevance of photoelectric charging of dust grains in the
accretion disks atmospheres can be derived from a simple evaluation of $\Phi_e^a$ and $\Phi _e^a$ at $z=5H$ for our fiducial dust grain of radius 10~nm.

On the one hand,  the collisional charging term, $\Phi _e^a$, can be derived using 
as fiducial density $n(R,5H)$ and deriving the thermal velocity of the gas from the 
temperature of the accretion disk T(R) in the $\alpha$ prescription. Henceforth,  
\begin{align}
\Phi _e^a = 8.3\times 10^{18} ~ \frac {\text{electrons}}{\text{cm}^2 \text{s}} \xi _e \chi_e 
\left(\frac {T}{10^3 ~ {\rm K}} \right) ^{1/2}
\left(1+1.67 Q \left( \frac{10^3 ~ {\rm K}}{T}  \right) \right)
\label{eqn:phiea}
\end{align}
where $\chi _e$ is the fraction of electrons in the atmosphere.

\begin{figure}
\includegraphics[width=8.5cm]{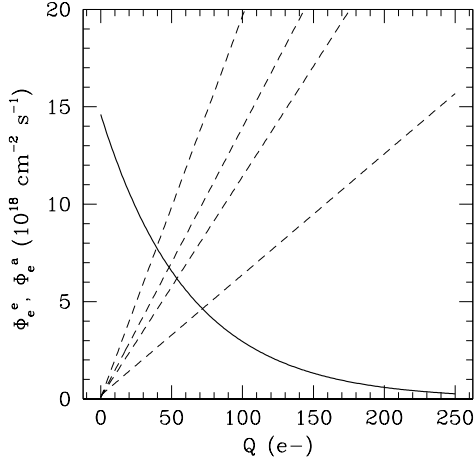}
\includegraphics[width=8.5cm]{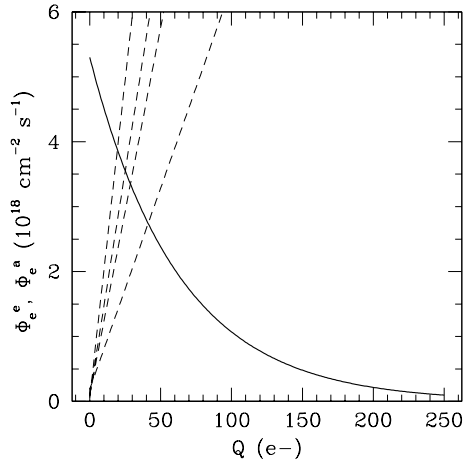}
\caption{
The photoelectric flux leaving a 10~nm dust grain irradiated by a TTS spectrum is shown as a function of the grain charge (thick line). This current produces a positive charging of the grains to a limit of 275 electrons. Electrons from the environment are attracted by the charged dust grain and an impinging electric current is created. This collisional current depends on the electric potential of the grain and on the thermal velocity
of the electrons. The collisional charging function is represented for various thermal velocities of the electrons corresponding to temperatures: 500, 1000, 1500, 5000 K with decreasing slopes. In the 
left panel, the photoelectric flux, $\Phi _e ^e$ is represented for $S_{19,0} = 0.073$. In the
the right panel,  the photoelectric flux, $\Phi _e ^e$ is represented for $S_{19,0}/e = 0.026$
}
\label{fig6}
\end{figure}

On the other hand, the photoelectric flux, $\Phi _e^e$, is given in terms of the 
stellar UV radiation flux entering the disk atmosphere, $S_{19,0}$ and the 
optical depth, $\tau$, as, 
\begin{align}
\label{eqn:phiee}
\Phi _e^e = 2 \times 10^{20} ~ \frac{\text{electrons}}{\text{cm}^2\text{s}} S_{19,0} \exp ^{-\tau }
\left( \frac {r_{grain}}{10{\rm ~ nm}} \right) ^{-1} 10^{-0.016 Q}
\end{align}
A simple illustration of the competition of both contributions, collisional charging 
versus photoelectric charge loses is shown in Fig.\ref{fig6} for a 10~nm radius dust grains. 
In the top panel, $\Phi _e^e$ has been evaluated at the inner radius of the dusty disk, 0.1~AU, and plotted as a function of the grain charge. The rapid exponential decay is clearly seen. In the 
bottom panel, the stellar radiation has been assumed to be reduced by a factor 1/e, as would correspond to  $z= 5H$. On top of that, the function of $\Phi _e^a (Q,T)$ has been plotted assuming in both cases that the ionization fraction is $\chi _e = 0.01$, and that 
the sticking coefficient is $\xi _e = 1$. $\Phi _e^a$ has been plotted for
T = 500, 1000, 1200, 1500 and 5000 K. The final charge of the dust grain is determined
by the intersection between $\Phi _e^e$ and the $\Phi _e^a (Q,T)$ corresponding to
the plasma temperature. As the temperature rises, the slope of $\Phi _e^a (Q,T)$
rises, and the equilibrium becomes at a larger charge  for the same number of photoionizing 
photons, $S_{19}$. If $S_{19}$ decreases, the region of the exponential  $\Phi _e^e$ curve
sampled by the same family of $\Phi _e^a (Q,T)$ curves, is at smaller
values of $Q$.

Notice that the temperature of the gas acts as a regulator of final dust grain charge
and determines the region dominated by the photoelectric charging, as much as the
photoionizing radiation flux.

\section{Conclusions}

The effect of the various UV radiation fields on dust grain charging has been analyzed by means of numerical simulations to show that TTSs spectra are more effective in dust grain charging than the often applied UV-background. The photoelectric yield has two main populations, a low energy 
population of electrons with $E= 2.5$ eV and a high energy population with $E=5.5-6$~eV; the energy dispersion within the populations is $\sim 1.3$~eV ($T\sim 1.5\times 10^4$~K). The high energy population is susceptible of dissociating the H$_2$  and ionizing some low ionization potential species, such as the Mg. These results add an additional role to dust on the chemistry of the layers just below the H$_2$ photoionization front. 

Finally a simple expression has been derived for the photoelectric currents from dust grains in TTSs environments that can be introduced on the modeling of protostellar disks and a simple example of the relevance of photoelectric charging is presented for a rather naive $\alpha$-disk model.

\section{Acknowledgements}
This research project was funded by an UCM - EEA Abel Extraordinary Chair Grant. AIG also acknowledges the support from the Spanish Ministry of Science and Innovation through grant:
AYA2008-06423-C03-01.

\section{Appendix}
Velocity Verlet, to start the simulation the initial position, $\mathbf{r}(t_{0})$, and velocity, $\mathbf{v}(t_{0})$, are required. Following must the acceleration, $\mathbf{a}(t)$, be determined for every time step to progress the system.
\begin{align}
{\mathbf{r}(t_{n+1}) = \mathbf{r}(t_{n})+\mathbf{v}(t_{n})\Delta t+\mathbf{a}(t_{n})\Delta t^{2}} \\
{\mathbf{v}(t_{n+1}) = \mathbf{v}(t_{n})+\frac{\mathbf{a}(t_{n})+\mathbf{a}(t_{n+1})}{2}}\Delta t
\end{align}

\noindent
(see appendix in \cite{swope82} for more details)

\newpage
\bibliographystyle{plain}
\bibliography{article_rev2}
\newpage

%

\end{document}